\documentstyle[prl,twocolumn,aps]{revtex}
\begin{document}
\input psfig
\draft

\title{Level statistics of quantum dots coupled to reservoirs}

\author{J\"urgen K\"onig$^1$, Yuval Gefen$^2$, and Gerd Sch\"on$^1$}

\address{$^1$Institut f\"ur Theoretische Festk\"orperphysik, Universit\"at
Karlsruhe, 76128 Karlsruhe, Germany\\
$^2$Department of Condensed Matter Physics, The Weizmann Institute of Science,
76100 Rehovot, Israel}

\date{\today}

\maketitle

\begin{abstract}
We study the effect of electron tunneling on the level statistics of quantum 
dots. While the coupling between individual levels and the electron
reservoir leads predominantly to the expected level broadening, the indirect
coupling of adjacent levels via 
the reservoir results in a new asymptotic level statistics and broadening.
These features, which are attributed to renormalized spectral weights rather 
than renormalized eigenvalues of the Hamiltonian, are observable in
the asymptotic frequency dependence in an absorption experiment.
\end{abstract}
 
{\it Introduction.} 
The statistics of the excitation spectra determines many physical
properties of quantum dots. 
These include thermodynamic  properties,  transport coefficients, 
the response to external a.c.\ fields, and statistical fluctuations 
\cite{Gor-Eli,Shk,Efe,Imr,Kam-Reu-Bou-Gef}.
For a model of non-interacting electrons confined within a moderately disordered 
quantum dot (diffusive disorder), the description of the single-particle levels 
for energies below the Thouless energy conforms to the Wigner-Dyson theory of 
random matrices (RMT) \cite{Efe,Efe-book,Bee,Guh-Mue-Wei}.
As a result of the level repulsion, on energy scales smaller than the average 
single-particle spacing $\Delta$, the statistics of the interlevel spacings for 
$s\ll 1$ is given by $P_o (s) = \pi s/2$, if the random system possesses time 
reversal symmetry, (orthogonal case) and by $P_u (s) = 32 s^2/\pi^2$ otherwise 
(unitary case) \cite{Meh}. 
Here $s$ is measured in units of $\Delta$.
This level statistics manifests itself in the response to low-frequency fields.
For instance, the a.c.\ dissipative conductance scales as the applied frequency 
$\omega_0$, or as $\omega_0^2$, for the two symmetry classes, respectively. 
This result, valid for the grandcanonical ensemble, is modified for the canonical
ensemble \cite{Gor-Eli,Shk,Kam-Reu-Bou-Gef,Kam-Gef,rem1}.

If the quantum dot with discrete bare levels is coupled to external degrees of 
freedom, these levels acquire a finite width $\Gamma$.
While the broadening is known to depend on the coupling and on the energy
measured from $\epsilon_F$ (see e.g. \cite{Zho-Spi-Tan-Alt}) it is commonly 
assumed that the levels statistics is unchanged. 
This means, the standard Kubo-Greenwood formalism can be employed, provided that 
the $\delta$-function describing the energy conservation is replaced by a 
Lorentzian with finite width.

In this Letter we study the influence of a dot-reservoir coupling on the 
effective level statistics and the asymptotic $\omega_0$-dependence of the 
absorbed power. 
A simple, though generic, model, where the quantum dot is coupled to an ideal 
reservoir of non-interacting electrons is used. 
In most of our analysis we ignore interactions in the dot. 
Breaking the RMT invariance by the choice of a preferred basis may modify the 
level statistics as it was demonstrated for disorder in metals in 
Ref.~\cite{Pic-Sha}.
In our case, the system is not invariant under rotation between the dot and the 
lead states.
We show that the coupling to the lead gives rise to a renormalization of both 
the low-frequency spectrum and the bare level broadening $\Gamma$. 
{\it We attribute the origin of the modified statistics and broadening
(Eqs.~(\ref{renormalized})-(\ref{P u})) to renormalized spectral weights rather 
than renormalized eigenvalues of the Hamiltonian.} 
Subsequently the frequency- and temperature-dependent absorption is calculated, 
(Eqs.~(\ref{power}) with (\ref{D inter o})-(\ref{D intra u})).
Finally we comment briefly on the effect of Coulomb interactions.

{\it Model.} The model Hamiltonian
\begin{equation}\label{hamiltonian}
  H = \sum_k \epsilon_k a_k^\dagger a_k 
  \!\! + \!\! \sum_{\sigma=1,2} \epsilon_\sigma c_\sigma^\dagger c_\sigma 
  + \sum_{k\sigma} \left[ T_{k\sigma} a_k^\dagger c_\sigma + h.c. \right]
\end{equation}
describes a quantum dot with two levels $\sigma =1,2$ coupled to the {\it same}
electron reservoir \cite{rem2}.
The tunnel matrix elements are assumed to be independent of the reservoir state, 
$T_{k \sigma} = T_{\sigma}$.
The Coulomb interaction will be discussed at the end this Letter.

Due to the coupling to the reservoirs, if both  dot levels are well separated, 
each level acquires a finite width related to 
$\Gamma_\sigma(\omega) = 2\pi \sum_k |T_{k\sigma}|^2\delta (\omega
-\epsilon_k)$.  However,
if the levels are degenerate, $\epsilon_1=\epsilon_2=\epsilon$, 
 a rotation within the dot subspace yields 
\begin{equation}
  H' = \sum_k \epsilon_k a_k^\dagger a_k 
  \! + \epsilon \!\! \sum_{\sigma=1,2} \! {c'}_\sigma^\dagger c'_\sigma 
  + \sum_k \left[ T_{k1} a_k^\dagger c'_1 + h.c. \right]
\end{equation}
where $ c_1' = (T_1 c_1+T_2 c_2)/\sqrt {|T_1|^2+|T_2|^2}$ and  
$ c_2' = (T_2 c_1-T_1 c_2)/\sqrt {|T_1|^2+|T_2|^2}$. In this form one
recognizes that only the new level 1 is coupled to a reservoir of
electrons (leading to a  
Lorentzian form of the spectral density) while the new level 2 is isolated 
(showing therefore a $\delta$-peaked spectral density).
Therefore, in the unrotated basis $c_1, c_2$, the spectral density of each level
consists of a Lorentzian part but also contains a $\delta$-function.
The sharp peak persists if the degeneracy is slightly lifted.

\begin{figure}
\centerline{\psfig{figure=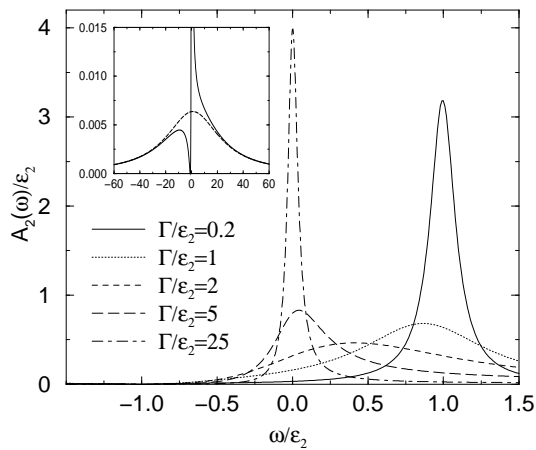,height=6.cm}}
\caption{Spectral density for level 2 with $\Gamma_1=\Gamma_2=\Gamma$ and 
$\epsilon_1 = -\epsilon_2 = 1$. 
For $\Gamma \ll \epsilon$ a peak with width $\tilde\Gamma \ll \Gamma$ evolves 
near zero.
The spectral density of level 1 is related to that of level 2 by
$A_1(\omega)=A_2(-\omega)$.
Inset: Spectral density for $\Gamma/\epsilon_2=25$ (solid line) showing a sharp 
peak sitting atop a Lorentzian with width $2\Gamma$ (dashed line).}
\label{fig1}
\end{figure}

{\it Spectral density.} In order to catch the evolution of the sharp peak in 
the spectral density quantitatively, we employ the equations of motion for the 
Green's functions
$G_{\sigma \sigma}(t):=-i\Theta (t) \langle \left\{ c_\sigma(t),c_\sigma(0) 
\right\} \rangle$.
Since the Hamiltonian Eq.~(\ref{hamiltonian}) is quadratic, we can find the exact
solution 
\begin{equation}
  G_{\sigma \sigma}(\omega) = \left[ \omega - \epsilon_\sigma - 
    \Sigma_{\sigma \sigma} - {\Sigma_{\sigma \bar\sigma} 
      \Sigma_{\bar\sigma \sigma} \over 
      \omega - \epsilon_{\bar\sigma} - \Sigma_{\bar\sigma \bar\sigma} } 
  \right]^{-1}
\end{equation}
with $\Sigma_{\sigma \sigma'}=\sum_k T_{k\sigma} T_{k\sigma'}^* / (\omega -
\epsilon_k)$, i.e. $\Sigma_{\sigma \sigma}(\omega+i0^+) = -i\Gamma_\sigma/2$ and
$\Sigma_{\sigma \bar\sigma}(\omega+i0^+)\Sigma_{\bar\sigma \sigma}(\omega+i0^+) 
= -\Gamma_\sigma \Gamma_{\bar\sigma} /4$
(for simplicity we put $\hbar=1$).
The spectral function follows from
$A_\sigma(\omega) = - {\rm Im} \, G_{\sigma \sigma} (\omega+i0^+)/\pi$.
For $\epsilon_1 \neq \epsilon_2$ it is
\begin{equation}\label{spec}
  A_\sigma(\omega)=
  {\Gamma_\sigma \left( \omega-\epsilon_{\bar\sigma} \right)^2 / 2\pi\over
    \left( \omega-\epsilon_\sigma \right)^2 
    \left( \omega-\epsilon_{\bar\sigma} \right)^2 
    + \left[ (\Gamma_1+\Gamma_2) \omega/4 - \hat\epsilon
      \right]^2 }
\end{equation}
with 
$\hat\epsilon = (\Gamma_1\epsilon_2 + \Gamma_2\epsilon_1)/(\Gamma_1+\Gamma_2)$.
In Fig.~\ref{fig1} we plot the spectral density of level 2 for several ratios 
$\Gamma/\Delta\epsilon$. For the following we define $\Delta \epsilon
= \epsilon_2-\epsilon_1$.  
If the bare levels are well separated, i.e., $\Delta\epsilon \gg \Gamma$, the 
spectral density is approximately a Lorentzian with width $\Gamma_\sigma$,
\begin{equation}
  A_\sigma(\omega) = {1\over \pi}{\Gamma_\sigma /2 \over 
    (\omega-\epsilon_\sigma)^2+(\Gamma_\sigma/2)^2},
\end{equation}
and the maximum is at the bare value $\epsilon_\sigma$, i.e., the level 
splitting remains unrenormalized $\Delta\tilde\epsilon = \Delta\epsilon$. 
We see, however, from Fig.~\ref{fig1} that things are different in the opposite
limit $\Delta\epsilon \ll \Gamma$.
Then, the level position is renormalized which leads to a modified level 
statistics, and the width is much smaller than the bare broadening.

For the degenerate case $\epsilon_1 = \epsilon_2$ we find
\begin{equation}
  A_\sigma(\omega)=
  {\Gamma_\sigma / 2\pi \over \left( \omega-\epsilon_\sigma \right)^2 
    + (\Gamma_1+\Gamma_2)^2/4 }
  + {\Gamma_{\bar\sigma}\delta(\omega - \epsilon_\sigma)\over \Gamma_1+\Gamma_2}
\end{equation}
and recover the expected $\delta$ peak.

{\it Modified level statistics.} 
For $\Delta\epsilon \ll \Gamma$, Eq.~(\ref{spec}) effectively reduces to the sum 
of two Lorentzians,
\begin{eqnarray}
  A_\sigma(\omega) &=& {\Gamma_\sigma\over \Gamma_1+\Gamma_2} {1\over \pi}
  {(\Gamma_1+\Gamma_2) /2 \over 
    (\omega-\epsilon_\sigma)^2+\left[(\Gamma_1+\Gamma_2)/2\right]^2}
  \nonumber \\
  &+& {\Gamma_{\bar\sigma}\over \Gamma_1+\Gamma_2} {1\over \pi}
  {\tilde\Gamma /2 \over (\omega-\tilde\epsilon_\sigma)^2+(\tilde\Gamma /2)^2}.
\end{eqnarray}
It shows a broad peak with width $\Gamma_1+\Gamma_2$, and a sharper one with 
width $\tilde\Gamma$ (see inset of Fig.~\ref{fig1}).
For the latter peak the level splitting $\Delta\tilde\epsilon$ and the width 
$\tilde\Gamma$ are given by
\begin{equation}\label{renormalized}
  {\Delta\tilde\epsilon \over \Delta\epsilon} = 
  {\tilde\Gamma \over \Gamma_1+\Gamma_2} = 
  {4\Gamma_1\Gamma_2\over (\Gamma_1+\Gamma_2)^4} (\Delta\epsilon)^2 \, .
\end{equation}

This result implies a new statistics for the 
renormalized interlevel spacing $\tilde s=\Delta\tilde\epsilon /\Delta$
in the regime $\tilde s \ll \Gamma/\Delta \ll 1$,
\begin{eqnarray}\label{P o}
  \tilde P_o (\tilde s) &=& {\pi \over 6} 
  \left[ {(\Gamma_1+\Gamma_2)^4\over 4\Gamma_1\Gamma_2\Delta^2} \right]^{2/3}
  \tilde s^{-1/3}
  \\ \label{P u}
  \tilde P_u(\tilde s) &=& {32 \over 3 \pi^2} 
  {(\Gamma_1+\Gamma_2)^4\over 4\Gamma_1\Gamma_2\Delta^2 } 
\end{eqnarray}
and a renormalized broadening 
$\tilde \Gamma = 4\Gamma_1\Gamma_2(\Delta\epsilon)^2/(\Gamma_1+\Gamma_2)^3$.
In the orthogonal case the distribution even diverges for 
$\tilde s\rightarrow 0$ with integrable divergence.

\begin{figure}
\centerline{\psfig{figure=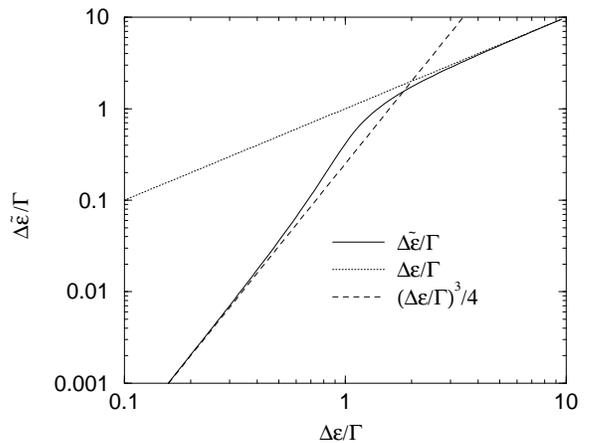,height=6.cm}}
\caption{Renormalized level spacing as a function of the bare one in double 
logarithmic plot for $\Gamma_1=\Gamma_2=\Gamma$.
The dotted and the dashed curve are the high- and low-energy 
(see Eq.~(\ref{renormalized})) asymptotes, respectively.}
\label{fig2}
\end{figure}

\begin{figure}
\centerline{\psfig{figure=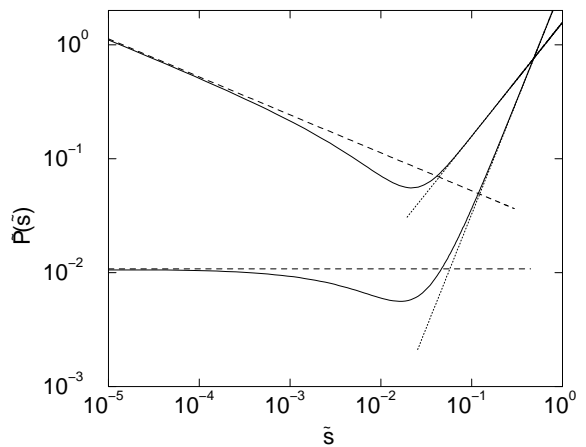,height=6.cm}}
\caption[a]{Solid lines: new level statistics $\tilde P (\tilde s)$ in the 
orthogonal (upper curve) and unitary ensemble (lower curve) with 
$\Gamma_1=\Gamma_2=\Gamma=\Delta/20$.
Dashed lines: low-energy asymptotes Eqs.~(\ref{P o}) and (\ref{P u}).
Dotted lines: high-energy asymptotes (Breit-Wigner distribution).
}
\label{fig3}
\end{figure}

Results for a wider range of energies are displayed in 
Figs.~\ref{fig2} and \ref{fig3} for the renormalized level spacing 
as a function of the bare one and for the new level statistics,
respectively, and compared to the low-energy asymptotic results
(\ref{renormalized}),  (\ref{P o}) and (\ref{P u}).

What is the origin of the  modified $\tilde P(\tilde s)$?
If we omit the tunneling part of the Hamiltonian Eq.~(\ref{hamiltonian}) the
the eigenvalues, $\epsilon_k$ and $\epsilon_\sigma$, have 
spectral weights totally within the corresponding states $k$ and 
$\sigma$.
It is tempting to assume that the coupling of the level $\sigma$ to the 
reservoir states $k$ leads to a renormalization of the eigenvalues 
$\epsilon_\sigma$, thus explaining the modified level statistics.
But this is not the case.
The crucial point is rather the renormalization of the eigenvectors.
Due to the tunneling each eigenvector acquires a finite overlap with
the bare dot states. 
What we see in Fig.~\ref{fig1} is the envelope of the square of this overlap.
The relevant physics is, therefore, a {\sl spectral weight
renormalization} effect rather than energy renormalization.

{\it Signatures in an absorption experiment.}
How can the evolution of the sharp peak be probed?
In the following we discuss consequences for an absorption experiment.
For a single quantum dot the absorption power of a photon with energy 
$\omega_0$, accompanied with a transition from level $\sigma$ to level
$\sigma'$,  
is given by
\begin{eqnarray}
  \int_{-\infty}^\infty d \, \omega \, && A_\sigma\left(\omega-{\omega_0\over 2}
  \right) A_{\sigma'}\left(\omega+{\omega_0\over 2}\right)\times
  \nonumber \\ &&
  f\left(\omega-{\omega_0\over 2}-\mu\right)
  \left[ 1 -  f\left(\omega+{\omega_0\over 2}-\mu\right) \right] \, .
\end{eqnarray}
Since in realistic experiments an ensemble of dots is probed, the total
signal $I_{\sigma\sigma'}$ is determined by the average over all possible 
configurations, i.e., we have to average over the chemical potential 
$\int_{\hat\epsilon-\Delta/2}^{\hat\epsilon+\Delta/2} {d\mu \over \Delta} \ldots
\approx \int_{-\infty}^\infty {d\mu \over \Delta} \ldots$
as well as over bare level separation $\int_0^1 ds \, P(s) \ldots$.
After performing the integral over $d\mu$ we get
\begin{eqnarray}\label{power}
  I_{\sigma\sigma'} &=& {\omega_0 \over \Delta} {e^{\beta \omega_0} \over
    e^{\beta \omega_0} - 1} D_{\sigma\sigma'}
  \\
  D_{\sigma\sigma'} &=& \int_0^1 ds \, P(s) B_{\sigma\sigma'}
  \\
  B_{\sigma\sigma'} &=& \int_{-\infty}^\infty d \, \omega 
  A_\sigma\left(\omega-{\omega_0\over 2}\right)
  A_{\sigma'}\left(\omega+{\omega_0\over 2}\right) .
\end{eqnarray}
The quantity $B_{\sigma \sigma'}$ probes the properties of the transition 
for a given level spacing.
It accounts for the renormalization of the level spacing and width.
The total signal, however, is determined by the average over all (bare) level 
spacings, $D_{\sigma \sigma'}$.
For $\sigma=\sigma'$ the absorption is accompanied by a transition within one 
level and for $\sigma\neq\sigma'$ a transition between the two levels.
In the following we are interested in the asymptotic behavior for 
$\omega_0 \rightarrow 0$, i.e., $\omega_0 \ll \Gamma \ll \Delta$.

In order to establish some reference frame we first approximate 
$A_\sigma(\omega)$ by (i) delta functions (which neglects the coupling of each 
level to the reservoir) and by (ii) Lorentzians with width $\Gamma_\sigma$ 
(which neglects the indirect coupling of the levels) before (iii) we use the 
exact form Eq.~(\ref{spec}).

(i) In the first case we find $D_{12} = P(\omega_0)$, i.e., 
$D_{12}^o = \pi^2 \omega_0/2$ and $D_{12}^u = 32\omega_0^2/\pi^2$,
while $D_{11} = D_{22} = D_{21} = 0$.
I.e.\ the absorption power vanishes for $\omega_0\rightarrow 0$.

(ii) In the second case
\begin{eqnarray}\label{B junior}
  B_{12} &=& {1\over 2\pi}{\Gamma_1+\Gamma_2 \over (\omega_0-\Delta\epsilon)^2
    + (\Gamma_1+\Gamma_2)^2/4}
  \\
  B_{21} &=& {1\over 2\pi}{\Gamma_1+\Gamma_2 \over (\omega_0+\Delta\epsilon)^2
    + (\Gamma_1+\Gamma_2)^2/4}
  \\
  B_{\sigma\sigma} &=& {1\over 2\pi}{\Gamma_\sigma \over \omega_0^2
    + \Gamma_\sigma^2},
\end{eqnarray}
which yields for $\omega_0=0$
\begin{eqnarray}\label{D junior o}
  D_{12}^o &=& D_{21}^o = {1\over 4} {\Gamma_1+\Gamma_2\over \Delta^2} 
  \ln {2\Delta\over \Gamma_1+\Gamma_2} 
  \\ \label{D junior u}
  D_{12}^u &=& D_{21}^u = {16\over \pi^3} {\Gamma_1+\Gamma_2\over \Delta^2} 
\end{eqnarray}
and $D_{\sigma\sigma}^o=1/ 4\Gamma_\sigma$, 
$D_{\sigma\sigma}^u=32/3\pi^3\Gamma_\sigma$.
Two points are remarkable.
First, due to the broadening there is always an overlap of the shifted spectral 
functions, i.e.\ even for $\omega_0=0$, $D_{12}$ remains finite.
Second, for systems with time reversal symmetry 
the $(\Delta\epsilon)^{-2}$ behavior of Eq.~(\ref{B junior}) leads
to logarithmic behavior, which is cut
off at low energies by the average level width $(\Gamma_1+\Gamma_2)/2$.

\begin{figure}
\centerline{\psfig{figure=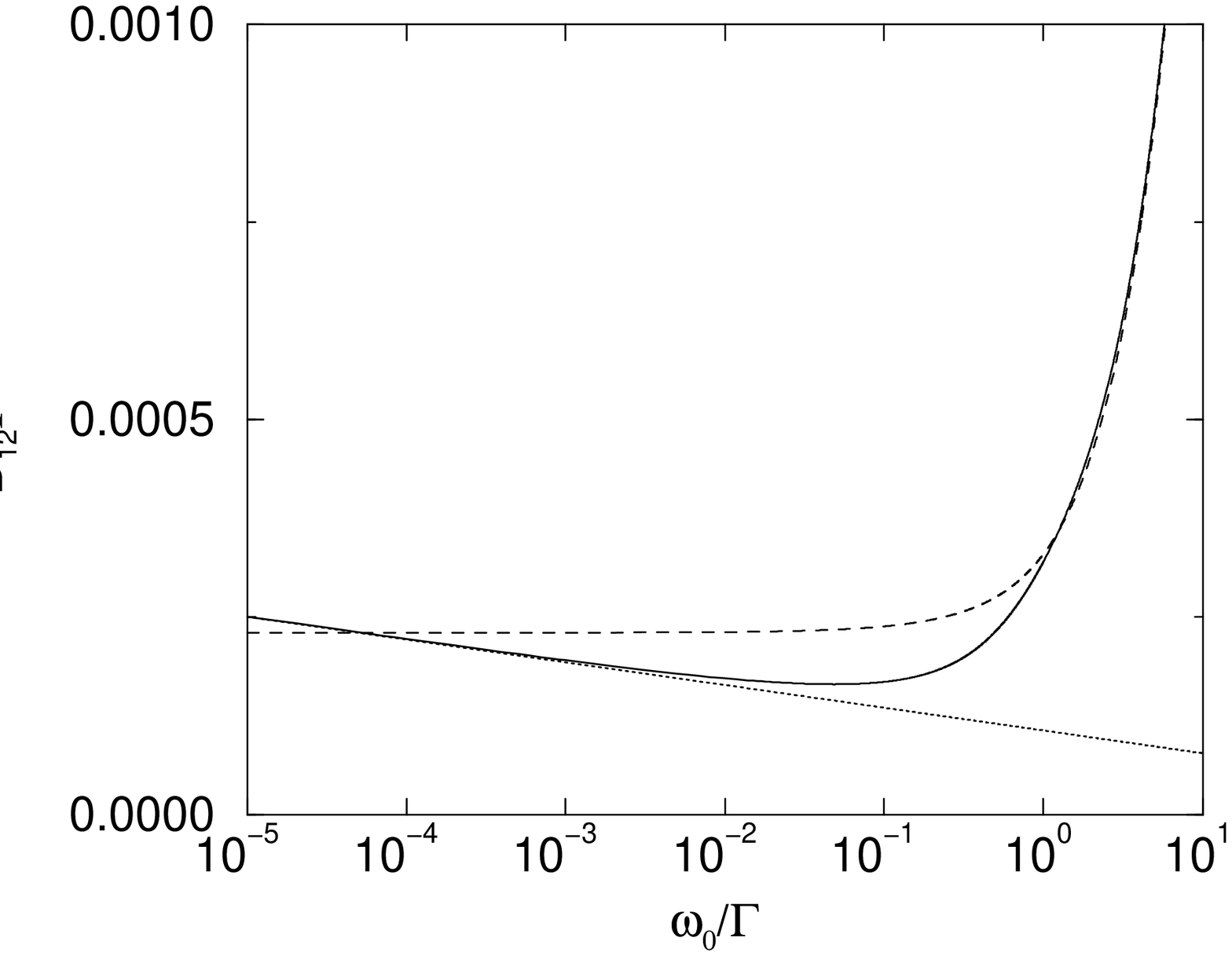,height=6.cm}}
\caption[a]{For the orthogonal ensemble the asymptotic absorption power is 
logarithmically divergent at small frequencies (solid line).
In the simplified model where the indirect coupling between the levels is
neglected, the low-energy behavior shows a saturation at $\Gamma$ (dashed line).
The dotted curve is the asymptote Eq.~(\ref{D inter o}).
}
\label{fig4}
\end{figure}

(iii) The renormalization of the level splitting and width due to indirect 
coupling of the levels to each other is fully included in the exact spectral 
function Eq.~(\ref{spec}). The leading terms are
\begin{eqnarray}\label{D inter o}
  D_{12}^o &=& D_{21}^o = {1 \over 8} {\Gamma_1+\Gamma_2\over \Delta^2} 
  \ln {\Delta\over \sqrt{\omega_0 {(\Gamma_1+\Gamma_2)^3 \over 
        4\Gamma_1\Gamma_2}}}
  \\ \label{D inter u}
   D_{12}^u &=&  D_{21}^u = {8 \over \pi^3} {\Gamma_1+\Gamma_2\over \Delta^2} 
\end{eqnarray}
and 
\begin{eqnarray}\label{D intra o}
  D_{\sigma\sigma}^o &=&{1 \over 4 \Gamma_\sigma}+ {1 \over 8} 
  {\Gamma_1+\Gamma_2\over \Delta^2}
  {\Gamma_{\bar \sigma} \over \Gamma_\sigma}  
  \ln {\Delta\over \sqrt{\omega_0 {(\Gamma_1+\Gamma_2)^3 \over 
        4\Gamma_1\Gamma_2}}}
  \\ \label{D intra u}
  D_{\sigma\sigma}^u &=&{32 \over 3\pi^3 \Gamma_\sigma}+ {16 \over \pi^3} 
  {\Gamma_1+\Gamma_2\over \Delta^2}
  {\Gamma_{\bar \sigma} \over \Gamma_\sigma}  \;.
\end{eqnarray}
We note that the low-energy cutoff 
$\sqrt{\omega_0 {(\Gamma_1+\Gamma_2)^3 \over 4\Gamma_1\Gamma_2}}$ is frequency
dependent and determines the asymptotic absorption power as can be seen
in Fig.~\ref{fig4}.
In comparison to Eqs.~(\ref{D junior o}) and (\ref{D junior u})
a factor $1/2$ arises due to details in the spectral density.

We remark here that for a symplectic ensemble we get similar result as for the 
unitary case Eq.~(\ref{D inter u}) and (\ref{D intra u}),
$D^s_{12}=D^s_{21}=(2^{16}/3^7\pi^4) (\Gamma_1+\Gamma_2)/\Delta^2$ and
$D^s_{\sigma \sigma}=2^{18}/(5 \cdot \,\,3^6 \pi^4 \Gamma_\sigma) + 
(2^{17}/3^7\pi^4) (\Gamma_{\bar\sigma}/\Gamma_\sigma)
(\Gamma_1+\Gamma_2) /\Delta^2$.

{\it Interaction.}
To get a qualitative understanding for the effect on interactions we include
a finite charging energy in 
the Hamiltonian $H \rightarrow H + U n_1 n_2$. As an example
 we consider the degenerate case $\Delta\epsilon =0$.
A rotation within the dot subspace, as discussed above, 
 can still decouple one new level from the reservoir.
But the levels still influence each other due to the charging energy term 
$U n_1' n_2'$ and the conclusion that there will be a delta-function peak in the 
spectral density is no longer valid.
If we neglect all terms in the equations of motion which would correspond to
correlations in the reservoirs we get a closed set of equations for the Green's 
functions. The resulting 
expression for $G_{\sigma\sigma}(\omega+i0^+)$ for arbitrary $U$ is too
lengthy to be presented here. However, for $U \rightarrow \infty$
 it simplifies to
\begin{equation}
  G_{\sigma\sigma}(\omega) = {1\over 2} \left(
    {1-\langle n \rangle \over \omega -\epsilon +i\Gamma} +
    {1-\langle n \rangle \over \omega -\epsilon - 2\hat\Sigma (\omega)} \right)
\end{equation} 
with ${\rm Im}\, \hat\Sigma (\omega+i0^+) = -\Gamma/2 f(\omega)$ and
${\rm Re}\,\hat\Sigma (\omega+i0^+) = \Gamma/2\pi \left[ \ln (\beta U / 2\pi) - 
{\rm Re}\, \Psi (1/2+i\beta\omega/2\pi) \right]$.
Here, $\Psi(z)$ is the digamma function and $f(\omega)$ is the Fermi function.
If the levels are far (in units of $\Gamma_\sigma$) from the Fermi level
they only contribute to the absorption power in the case 
$|\omega-\epsilon| \lesssim k_B T$ since otherwise the levels are totally empty 
or filled.
Then, the second term acquires a width of the order of $\Gamma$ and does not show
a sharp peak.
If the levels are close to the Fermi level also the low-temperature regime 
is important.
But then the spectral density shows a richer structure related to collective 
many-particle states. 
Similar effects lead to Kondo physics in a single-level spin-degenerate quantum
dots with spin conservation (see e.g. \cite{kss,ksss}).
The analysis of the level's properties and their effect on the absorption power 
becomes more complicated and is out of the aim of this Letter.

In summary, our model analysis produced a modified statistics of the broadened 
levels due to a renormalization of the spectral weights.
As a consequence, in the orthogonal case the asymptotic absorption power depends 
logarithmically on the external frequency.

{\it Acknowledgments.}
We acknowledge useful discussions with J.L. Pichard. 
J.K. and Y.G. acknowledge the hospitality of the Weizmann Institute of Science 
and the University of Karlsruhe, respectively. 
This work was supported by the DFG as part of 
SFB 195, by the German-Israeli Foundation (GIF), by the Israel Science
Foundation founded by the Israel Academy of Sciences and Humanities - 
Centers of Excellence Program, and by the U.S.-Israel Binational-Science 
Foundation (BSF).

\begin{references}
\bibitem{rem1} 
This applies for the limit of $\Delta \gg \omega_0 >\Gamma$, where $\Gamma$ is 
the level's broadening.
\bibitem{rem2} 
This describes either the situation in which the spin labeled by $\sigma$ is not
a conserved quantum number or in which $\sigma$ labels some other non-conserved
quantum number.
In the latter case spin is ignored either due to strong Zeeman splitting 
or since for a non-interacting quantum dot the total Hamiltonian separates into 
two parts describing either spin.
The corresponding levels do not influence each other.
\end{references}
\end{document}